\documentclass[12pt]{article}%
\usepackage{amsmath}
\usepackage{graphicx}%
\usepackage{amsfonts}%
\usepackage{amssymb}

\begin{document}

\author{C. S. Unnikrishnan\thanks{E-mail address: unni@tifr.res.in}\\\textit{Gravitation Group, Tata Institute of Fundamental Research,}\\\textit{Homi Bhabha Road, Mumbai - 400 005, India \&}\\\textit{NAPP Group, Indian Institute of Astrophysics,}\\\textit{Koramangala, Bangalore - }560 034, India}
\title{Casimir Energy Density at Planck Time: Cosmic Coincidence or Double Solution
to the Cosmological Dark Energy Problem?}
\date{July 2002}
\maketitle

\begin{abstract}
The Casimir energy density calculated for a spherical shell of radius equal to
the size of the Universe projected back to the Planck time is almost equal to
the present day critical density. Is it just a coincidence, or is it a
solution to the `cosmic dark energy' and the `cosmic coincidence' problems?
The correspondence is too close to be ignored as a coincidence, especially
since this solution fits the conceptual and numerical ideas about the dark
energy, and also answers why this energy is starting to dominate at the
present era in the evolution of the Universe.

\end{abstract}

It is startling to notice that the Casimir energy density of a spherical
bounded space with its radius equal to the size of our present Universe scaled
back to its size at the Planck time is almost exactly the critical energy
density. It is perhaps not reasonable to discard this as a coincidence, since
it solves the two important current problems in cosmology with vacuum energy
\cite{iap-dark}, namely the problem of the smallness of the cosmological
vacuum energy density and the problem of cosmic coincidence of the near
equality of the vacuum energy density and the \emph{present }matter density.

There have been several calculations of the Casimir energy of a conducting
spherical shell bounding three dimensional space \cite{miloni,milton}. The
result for the electromagnetic vacuum inside a shell of radius $R$ is
\begin{equation}
\rho_{C}=\frac{0.046\hbar c}{(4\pi/3)R^{4}}%
\end{equation}

For calculations with scalar field one gets a similar expression with a
numerical coefficient different by a factor of order unity (A factor of 2
comes from polarization degrees of freedom) \cite{milton}. An estimate of such
a vacuum energy density for the present Universe with the Hubble radius of the
order of $10^{28}$ cm gives a value, $\rho_{H}$, negligible by a factor
$10^{123}$ compared to the present day critical density of approximately
$\rho_{0}=1.9\times10^{-29}$ g/cm$^{3}$. On the other hand, an estimate using
Eq. 1 for a Planck size Universe gives a value of the order of the Planck
energy density, $\rho_{P}$, $3\times10^{92}$ g/cm$^{3}$ -- enormous compared
to $\rho_{0}.$ The fact that the number we need to fit the present
observations, the critical density itself, is the geometrical mean of
$\rho_{H}$ and $\rho_{P}$ might be a genuine clue, or might just be a coincidence.

\textit{However, the Planck length and the Hubble length are not the
physically relevant scales for an estimate of the vacuum energy in the
Universe}. It is well known that a Universe with Planck size at Planck time
could not have evolved into the present Universe without large inflationary
factors \cite{zeldo}. An important and physically relevant scale for the
quantum cosmology of an expanding Universe would be its \ ``total size'' at
Planck time. This is much larger than the Hubble scale at Planck time, since
the horizon scale and Hubble scale grows linearly with time whereas the
typical scale factor grows as only $t^{1/2}$ and $t^{2/3}$ during the
radiation dominated and matter dominated eras respectively. Therefore, the
horizon scale and the size of the `visible Universe' (part of the Universe
that can be in causal contact at some time) were much smaller than the total
extent of the Universe at Planck time.

We do not know the total extent of the Universe. It could be infinite, but it
could be just very large and finite. All we can say with definiteness now is
that it is certainly larger than or equal to about $10^{28}$ cm. Extrapolating
backwards in evolution to the Planck time from the present Hubble scale of the
order of $10^{28}$ cm gives, in the standard big bang picture with a critical
evolution, the `diameter' of our presently observable Universe projected to
Planck time \cite{kolb} as
\begin{equation}
D_{Pl}\simeq1.4h^{-1}10^{-3}~cm
\end{equation}

With the observed value of the Hubble parameter, $h\simeq0.7,$ we get
$D_{Pl}\simeq20$ microns. Therefore the size of the entire Universe at Planck
time was larger than 20 microns, and could be a few times larger if the
Universe is finite, say $20-100$ $\mu m$.

Now I estimate the radius of the bounded space that will generate a Casimir
energy density that is equal to $2/3$ the critical energy density at present,
$(2/3)\rho_{0}=1.3\times10^{-29}$ g/cm$^{3}$, which is the estimated amount of
dark energy in the Universe at present. From eq. 1 for the Casimir energy
density (and dividing by $c^{2}$ to get the mass density),
\begin{equation}
D=2R=2\left(  \frac{0.046\hbar}{(4\pi/3)(2/3)\rho_{0}c}\right)  ^{1/4}%
\simeq55\mu m
\end{equation}
With a scalar field this number is about $46$ $\mu m.$ The Casimir energy of
the electromagnetic field is not the relevant energy here, since it does not
have the required equation of state $p=-\rho.$ With the scalar field, there is
the possibility to get the Casimir energy from its quantum fluctuations with
the required equation of state. Therefore, \emph{the size of the Universe that
contains a Casimir energy density equal to the critical density is in the same
range as the present Hubble size of the Universe extrapolated back to the
Planck time}. This is surprisingly good agreement. If this number was less
than $10$ $\mu m$ the hypothesis that the present dark energy could be the
Casimir energy generated at Planck time would have been ruled out immediately
since the present Hubble size is known well within a factor of 2.

It is possible that this is a mere coincidence (Then, contrary to Einstein's
assertion, G is malicious!). But, here the correspondence is too close to be
ignored as a mere coincidence, especially since there is no accepted solution
to the various questions raised by the possible presence of a small vacuum
energy density comparable to the critical energy density in the Universe. In
fact, a more precise estimate of the scalar field Casimir energy density for
the Universe, with its size at Planck time taken as the present horizon scale
projected to the Planck time, might show that $\rho_{C}\approx\rho_{0}$ even
more closely than we have estimated. There is a calculation of the scalar
field vacuum energy density for the manifold $M^{4}\times S^{3},$ product of
the 4-d Minkowski space-time and a compact 3-d space, giving a similar
estimate, with the energy density becoming comparable to the critical energy
density for $D\simeq20$ microns \cite{milton}. In our scenario, a size of the
order of 20 microns is not the size of the compact extra dimension, but\ it is
the size of the entire bounded Universe at Planck time.

It may be noted that \emph{the estimated size of the Universe at Planck time
is the geometric mean of the Planck length and the present Hubble length}.
This is a coincidence without any physical significance, since this relation
will change as a function of time. Thus we are able to explain why the present
critical density is approximately the geometric mean of the Planck energy
density and the vacuum energy density calculated using Hubble length as the
relevant scale. While this analysis suggests that there is no special physical
significance to this fact, the considerations in ref.\cite{ugr} clarify the
possible relations between these length scales.

The scenario I described naturally answers the important unresolved question
why the dark energy density has started to dominate the matter energy density
only recently in the evolution of the Universe. This is the Cosmic Coincidence
problem: why is the vacuum energy density comparable to the \emph{present day}
critical density? Due to the special equation of state, $p=-\rho,$ the vacuum
energy density remains constant as the Universe expands. During the early
evolution of the Universe an energy density of $1\times10^{-29}$ g/cm$^{3}$ in
the Casimir form is totally insignificant compared to the energy density in
radiation or in matter. As the Universe expanded the energy density in
radiation and matter dropped as $R^{4}$ and $R^{3}$ respectively. The scalar
vacuum energy density remained constant and insignificant till recently when
the matter energy density has dropped to about $10^{-29}$ g/cm$^{3}$, and the
vacuum energy density has just exceeded the matter density. The magnitude of
the Casimir energy at Planck time for a Universe of size of approximately
40-50 microns is such that it will dominate matter density after about 14
billion years, if the evolution of the Universe is as in the standard big bang picture.

Thus we have the double solution cosmologists are seeking -- explanation of
the present value of the vacuum energy density and an explanation for the
question why the vacuum energy density has started dominating \emph{now}.

This scenario addresses the cosmological constant problem directly; the
question why we do not see any effect of the infinite zero point energy of the
quantum vacuum. The cosmological constant is small because \emph{quantum
vacuum has no energy density that can act as a source of gravity}, and only
the Casimir type vacuum energy density arising from bounding the vacuum in
finite sized boundaries has any physical relevance. \emph{Free quantum vacuum
is truly empty}. This is of course restating Schwinger's view on the vacuum
energy density, and I think that it is an important stand that can solve many
problems. At present, such an idea is consistent with all experimental and
theoretical facts.

In conclusion, I wish to stress a viewpoint that we are probably faced with
several observational evidences that are indicative of a quantum origin of our
Universe -- evidence we do not take seriously today, but might be compelled to
accept tomorrow. Earlier we have pointed out that some of the standard
``classical'' observations are perhaps evidence for quantum cosmology
\cite{ugr}. In this paper I have pointed out a possible solution to the
cosmological dark energy problem in terms of the Casimir energy of the entire
bounded Universe at Planck time. The numerical estimates match the present
observational parameters well, and it also answers the query why the dark
energy has started dominating relatively recently.\medskip

\noindent\textbf{Acknowledgments}: The collaboration with G. T. Gillies and R.
C. Ritter on an earlier paper was important for initiating the thoughts
presented here. Also important were the several review talks at the 18th IAP
astrophysics colloquium, 2002, in Paris, which highlighted the desperate urge
for some theoretical clue to handle the accumulating observational evidence
supporting the presence of dark energy.

\end{document}